\documentclass[%
reprint, superscriptaddress,
 amsmath,amssymb,
 Aps, 
prd,
longbibliography,
]{revtex4-1}
\AtBeginDocument{%
    \newwrite\bibnotes
    \def\bibnotesext{Notes.bib}
    \immediate\openout\bibnotes=\jobname\bibnotesext
    \immediate\write\bibnotes{@CONTROL{REVTEX41Control}}
    \immediate\write\bibnotes{@CONTROL{%
    apsrev41Control,author="08",editor="1",pages="1",title="1",year="1"}}
     \if@filesw
     \immediate\write\@auxout{\string\citation{apsrev41Control}}%
    \fi
}%
\usepackage{graphicx}
\usepackage{dcolumn}
\usepackage{bm}

\begin{document}

\preprint{APS/123-QED}

\title{Gate-controlled supercurrent in epitaxial Al/InAs nanowires}

\author{Tosson Elalaily}\thanks{Contributed equally to this work}
 \affiliation{Department of Physics, Budapest University of Technology and Economics and Nanoelectronics 'Momentum' Research Group of the Hungarian Academy of Sciences, Budafoki ut 8, 1111 Budapest, Hungary\\}
\affiliation{Department of Physics , Faculty of Science, Tanta University, Al-Geish St., 31527 Tanta, Gharbia, Egypt.\\}

\author{Oliv\'er K\"urt\"ossy} \thanks{Contributed equally to this work}
 \affiliation{Department of Physics, Budapest University of Technology and Economics and Nanoelectronics 'Momentum' Research Group of the Hungarian Academy of Sciences, Budafoki ut 8, 1111 Budapest, Hungary\\}

\author{Zolt\'an Scher\"ubl}
 \affiliation{Department of Physics, Budapest University of Technology and Economics and Nanoelectronics 'Momentum' Research Group of the Hungarian Academy of Sciences, Budafoki ut 8, 1111 Budapest, Hungary\\}
 \affiliation{Univ. Grenoble Alpes, CEA, Grenoble INP, IRIG, PHELIQS, 38000 Grenoble, France\\}
 
\author{Martin Berke}
 \affiliation{Department of Physics, Budapest University of Technology and Economics and Nanoelectronics 'Momentum' Research Group of the Hungarian Academy of Sciences, Budafoki ut 8, 1111 Budapest, Hungary\\}
 
 \author{Gerg\H{o} F\"ul\"op}
 \affiliation{Department of Physics, Budapest University of Technology and Economics and Nanoelectronics 'Momentum' Research Group of the Hungarian Academy of Sciences, Budafoki ut 8, 1111 Budapest, Hungary\\}

\author{Istv\'an Endre Luk\'acs}
 \affiliation{Center for Energy Research, Institute of Technical Physics and Material Science, Konkoly-Thege Mikl\'os \'ut 29-33., H-1121, Budapest, Hungary\\}

\author{Thomas Kanne}
 \affiliation{Center for Quantum Devices and Nano-Science Center, Niels Bohr Institute, University of Copenhagen, Universitetsparken 5, DK-2100, Copenhagen, Denmark\\}
 
\author{Jesper Nygård}
 \affiliation{Center for Quantum Devices and Nano-Science Center, Niels Bohr Institute, University of Copenhagen, Universitetsparken 5, DK-2100, Copenhagen, Denmark\\}

\author{Kenji Watanabe}
 \affiliation{Research Center for Functional Materials, National Institute for Material Science, 1-1 Namiki, Tsukuba, 305-0044, Japan\\}
 
 \author{Takashi Taniguchi}
 \affiliation{International Center for Materials Nanoarchitectonics, National Institute for Materials Science, 1-1 Namiki, Tsukuba 305-0044, Japan\\}
 
\author{P\'eter Makk}
 \email{makk.peter@ttk.bme.hu}
 \affiliation{Department of Physics, Budapest University of Technology and Economics and Nanoelectronics 'Momentum' Research Group of the Hungarian Academy of Sciences, Budafoki ut 8, 1111 Budapest, Hungary\\}
 
 \author{Szabolcs Csonka}
 \email{szabolcs.csonka@mono.eik.bme.hu}
 \affiliation{Department of Physics, Budapest University of Technology and Economics and Nanoelectronics 'Momentum' Research Group of the Hungarian Academy of Sciences, Budafoki ut 8, 1111 Budapest, Hungary\\}

\begin{abstract}

Gate-controlled supercurrent (GCS) in superconductor nanobridges has recently attracted attention as a means to create superconducting field effect transistors. Despite the clear advantage for applications with low power consumption and high switching speeds, the microscopic mechanism of the field effect is still under debate. In this work, we realize GCS for the first time in an epitaxial superconductor, which is created as a shell on an InAs nanowire. We show that the supercurrent in the epitaxial Al layer can be switched to the normal state by applying $\simeq\pm$ 23$\,$V on a bottom gate insulated from the nanowire by a crystalline hBN layer. Our extensive study on the temperature and magnetic field dependencies of GCS suggests that hot electron injection alone cannot explain our experimental findings.

\end{abstract}

\maketitle


\section*{Introduction}
Superconductor circuits have become promising building blocks in various architectures for quantum computing devices \cite{krantz2019quantum,devoret2004superconducting}, single photon detectors \cite{wang2009superconducting,natarajan2012superconducting}, quantum-limited amplifiers (parametric amplifiers) \cite{aumentado2020superconducting}, phase-coherent caloritronics \cite{giazotto2012josephson,fornieri2017towards}, ultra-sensitive magnetometers \cite{luomahaara2014kinetic,cleuziou2006carbon} and fast classical supercomputers \cite{frasca2019hybrid,soloviev2017beyond}. In the latter, the superconducting electronics are integrated with semiconductor technology. In particular, rapid single flux quantum (RSFQ) devices become more desirable due to their fast switching speed and low power consumption \cite{soloviev2017beyond}. Since RSFQ consists of a superconducting loop with Josephson junctions along with on-chip coils, its up-scaling remains a challenge. In order to realize a scalable network, electrical control via gate electrodes would be desirable. 

Very recently a striking new effect was observed in metallic nanostructures. In superconductor nanobridges the supercurrent can be controlled by applying a voltage on a closely spaced gate electrode \cite{de2018metallic,de2019josephson,paolucci2019field,de2020niobium,puglia2020electrostatic,puglia2020vanadium,rocci2020gate,paolucci2019magnetotransport,paolucci2019field2,puglia2021gate,alegria2021high, ritter2021superconducting,golokolenov2021origin}. By increasing the gate voltage beyond a certain threshold, the supercurrent is monotonically decreased until switching of the device from the superconducting to normal state. Previous works show GCS in thin metallic nanowires \cite{de2018metallic,rocci2020gate,alegria2021high,ritter2021superconducting}, in proximitized normal metal in superconductor-normal-superconductor (SNS) junction \cite{de2019josephson} and in Dayem nanobridges \cite{de2020niobium,paolucci2019magnetotransport,puglia2020vanadium}. In addition to high speed superconductor switches \cite{ritter2021superconducting}, observation of GCS in superconductor nanostructures led to realization of new superconductor nanodevices such as phase shifter \cite{paolucci2019field2} and gate-controlled superconductor half-wave nanorectifier \cite{puglia2020vanadium}. On the other hand, the inverse of this effect was observed in type II superconductors, where the supercurrent is enhanced by bipolar increase of the gate voltage \cite{rocci2020large}. Since GCS is bipolar with gate voltage, the origin of the effect being unclear. Some works explained the suppression of supercurrent by electric field induced perturbation of the superconducting state \cite{mercaldo2020spectroscopic,mercaldo2020electrically,chirolli2020impact}, like the superconductor Swinger effect \cite{solinas2021sauter}, others attributed the effect to injection of high energy quasiparticles tunneling from the gate electrodes \cite{ritter2021superconducting,alegria2021high,golokolenov2021origin}. The injected high energy electrons create a large number of quasiparticles and suppress the superconducting state. Despite the physical origin of the effect is unclear, gate-controlled nanobridges can work at ultra low power and highly switching speed. In addition, their configuration can be easily scaled up which provides a promising building block for superconductor switches in modern architectures of supercomputers and on-chip circuits for quantum computers.

Recently InAs semiconductor nanowires with epitaxial Al superconducting shells \cite{krogstrup2015epitaxy}, have become the primary platform for research on various potential quantum bit devices. Concepts have been developed for topologically protected qubits \cite{prada2020andreev} and even their surface code \cite{plugge2016roadmap} based on hybrid nanowires and they are also promising for realization of Gatemon or Andreev qubits  \cite{hays2018direct,larsen2015semiconductor}.  In all these quantum hardwares the gate tunable superconductivity would be highly desirable allowing an additional experimental control knob of the system. However, up to now, all gate tuning experiments were performed on polycrystalline materials and therefore it has not been clear if GCS exists for these highly crystalline materials.
In this work, we studied for the first time the superconducting gating effect in single crystalline Al shells grown on InAs nanowires. We will show in the following that the superconducting state can be switched off by application of a gate voltage on an epitaxial Al/InAs nanowire and provide detailed characteristics for the gating behaviour.
\section*{Experiments}
\subsection{Device outline}
We have used  InAs nanowires grown by molecular beam epitaxy (MBE) with a length of 5.5 $\mu$m using gold nanoparticles as catalysts. After InAs nanowire growth, Al shell layer of thickness $20\,$nm was epitaxially grown by deposition within the MBE chamber at low temperature. By rotating the substrate during Al growth, the Al shell layer is grown in all InAs nanowire facets resulting in fully covered nanowires \cite{krogstrup2015epitaxy}. 

\begin{figure}[tb!]
	\includegraphics[width=\columnwidth]{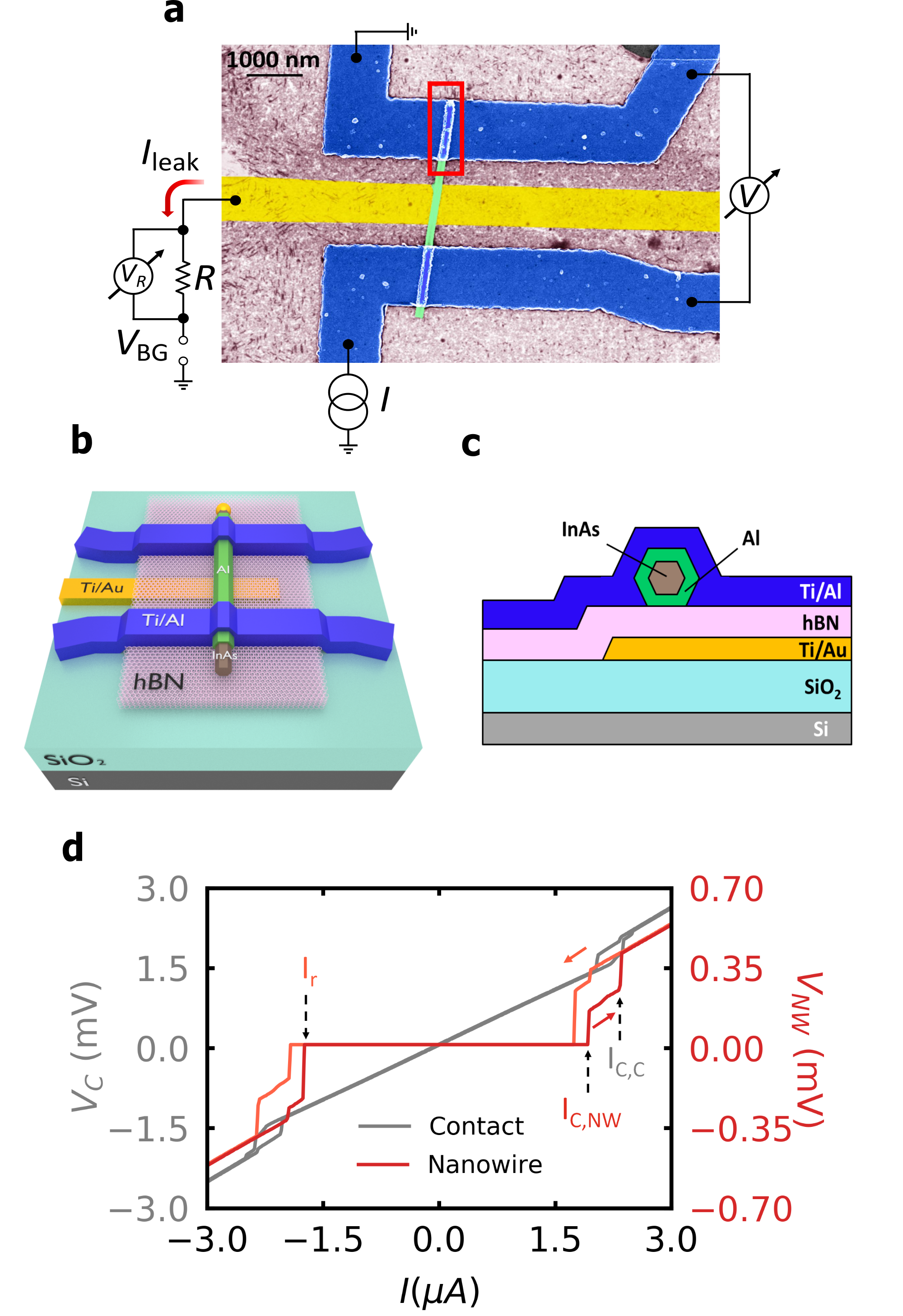}
	\caption{\label{fig:device}\textbf{Device configuration}. \textbf{a} False colored SEM image of the fabricated device with schematics of the device circuit. \textbf{b} Schematic of the device with $45^{\circ}$ angle view and side view in \textbf{c}. \textbf{d} (I-V) characteristics of the nanowire device (red curve) at 40 mK with two different switchings at $I_{\mathrm{C,NW}}$ and $I_{\mathrm{C,C}}$ in case of the bias current is ramped from negative to positive values (red arrow). In the opposite ramping direction (orange arrow), It switches back at the retrapping current $I_{\mathrm{r}}$. The measurement of pair of Al contacts (grey curve) has switching at the same value of $I_{\mathrm{C,C}}$.     }
\end{figure}

 To investigate the gate-controlled supercurrent in the epitaxial Al layer, nanowire-based device was fabricated as shown in Fig.~\ref{fig:device}a-c. A metallic gate from Ti/Au (yellow colored) with thicknesses of 7/33 nm was fabricated on an intrinsic Si wafer with a 290 nm thick oxide layer. To insulate the gate from the wire, 20-30 nm thick hBN (pink colored) was stacked on the gate electrode with PDMS-based dry transfer technique. hBN provides an excellent single crystal insulator between gate and wire serving as a tunnel barrier \cite{britnell2012electron,bretheau2017tunnelling,fu2014large,kamalakar2014enhanced}. The nanowire (gray) with Al shell (green) was deposited by a micromanipulator on top of the hBN layer. Two pairs of Al contacts (blue colored) have been fabricated on the top of the nanowire with a distance of 1.5  $\mu$m to allow four-terminal measurements for the nanowire. More details about fabrication are given in Methods section.   
 
 The current-voltage (I-V) characteristics of the nanowire device measured at 40 mK clearly show a well developed zero resistance state (see red curve in Fig.~\ref{fig:device}d) corresponding to a supercurrent flowing through the Al shell of the wire. Two clear switches to a finite resistance state are observed at $I_{\mathrm{C,NW}}$ = 1.94 $\mu$A and $I_{\mathrm{C,C}}$ = 2.34 $\mu$A. Similar multiple transitions were observed in suspended Ti nanowire \cite{rocci2020gate}. The nanowire device shows a hysteretic behaviour and switches back at the retrapping current $I_{\mathrm{r}}$ = 1.74 $\mu$A when the measurements were carried out in the opposite ramping direction. To identify the origin of both $I_{\mathrm{C,NW}}$ and $I_{\mathrm{C,C}}$, we have separately measured (I-V) curves for each horizontal pairs of contacts (blue electrodes in Fig.~\ref{fig:device}a) using 2-probe method. The measurements for one of the pairs with grey curve have switching current value at $I_{\mathrm{C,C}}$ value, while the other electrode switches at 7.5 $\mu$A (see supporting information). From this, we could attribute $I_{\mathrm{C,NW}}$ to the switching current of the nanowire segment while $I_{\mathrm{C,C}}$ to the switching current of the small metallic electrode segment (blue) above the nanowire marked by the red rectangle in Fig.~\ref{fig:device}a.
 
\subsection{Gate voltage dependence}

\begin{figure}[tb!]
	\includegraphics[width=\columnwidth]{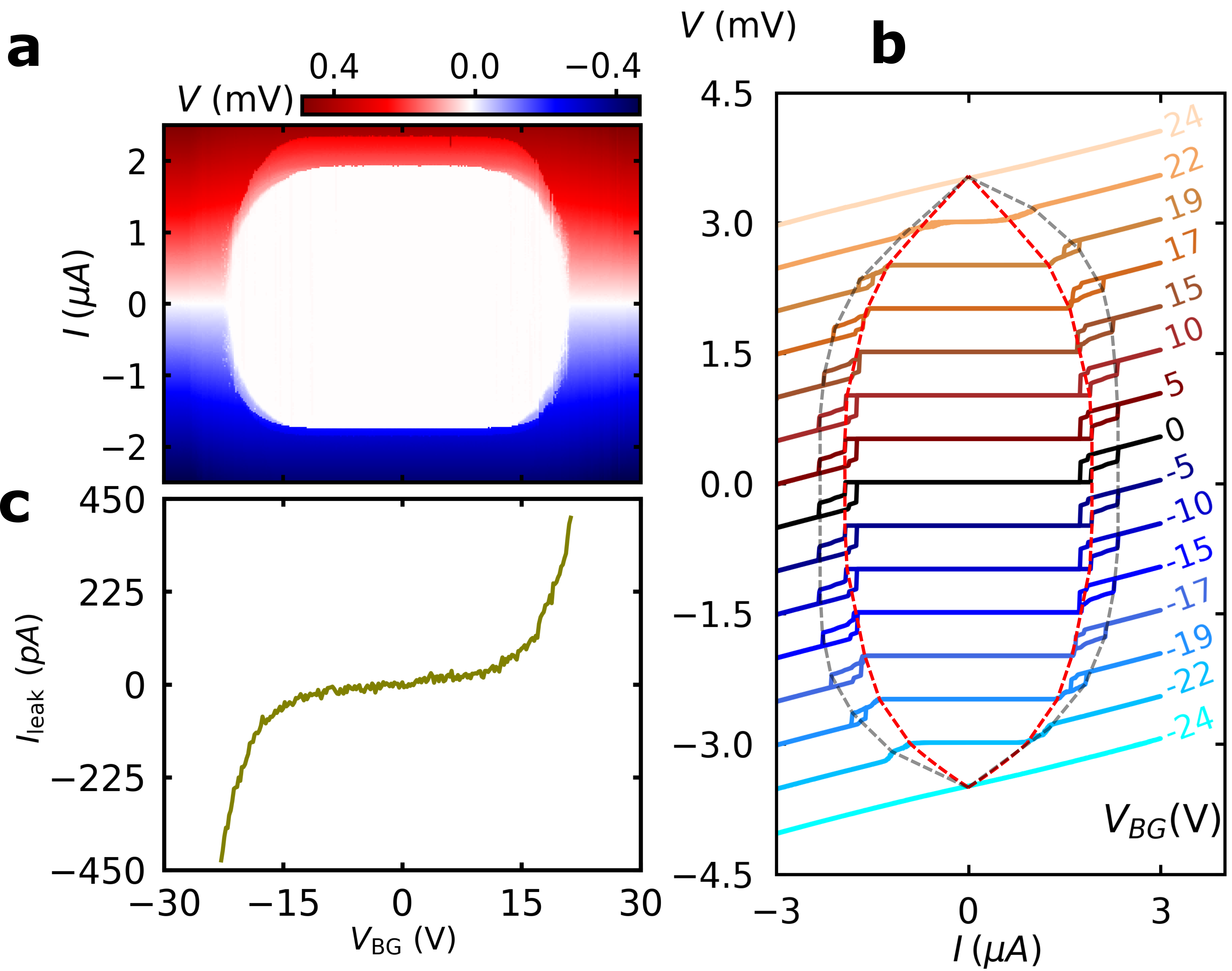}
	\caption{\label{fig:gatedepend}\textbf{Gating of supercurrent} \textbf{a} (I-V) characteristics of the nanowire device as a function of bipolar voltage applied to the bottom gate $V_{\mathrm{BG}}$ as I was sweeping from negative to positive values. \textbf{b} High resolution (I-V) curves  measured at selected gate voltages and separated on y-axis for better visibility. The red and gray dashed lines traces the suppression of $I_{\mathrm{C,NW}}$ and $I_{\mathrm{C,C}}$ with increasing $V_{\mathrm{BG}}$. \textbf{c} The measured and corrected leakage current from bottom gate to nanowire device as a function of $V_{\mathrm{BG}}$.}
\end{figure}

 The dependence of the supercurrent on the gate voltage was investigated by measuring the (I-V) curve of the nanowire device as a function of the bottom gate voltage, $V_{\mathrm{BG}}$ (see Fig.~\ref{fig:gatedepend}a). The white regions represent the zero resistance state. By increasing $V_{\mathrm{BG}}$ with either negative or positive polarity, both $I_{\mathrm{C,NW}}$ and $I_{\mathrm{C,C}}$ remains constant until  $V_{\mathrm{BG}} \simeq \pm 12 \,$V, then both are suppressed together up to full suppression at pinch-off gate voltage $V_{\mathrm{BG0}} \simeq \pm 23\,$V where the device is switched to normal state. The fine characteristics of the I-V curves are better visible in Fig.~\ref{fig:gatedepend}b, where the red and gray dashed lines trace the suppression of $I_{\mathrm{C,NW}}$ and $I_{\mathrm{C,C}}$ with increasing $V_{\mathrm{BG}}$, respectively. We should note that with increasing $V_{\mathrm{BG}}$, the difference between the retrapping current $I_{\mathrm{r}}$ and $I_{\mathrm{C,C}}$ is decreased and fully vanished at $V_{\mathrm{BG}} \simeq \pm 22 \,$V.

 The leakage current $I_{\mathrm{leak}}$ from the gate to the nanowire device was measured within $\pm V_{\mathrm{BG0}}$ window by recording voltage across the preresistor on the gate (see Fig.~\ref{fig:device}a). The leakage current was then corrected by subtracting the leakage between lines of the cryostat (see supporting information) \cite{ritter2021superconducting}. 
 The leakage current shows an exponential increase with $V_{\mathrm{BG}}$ for both polarities (see Fig.~\ref{fig:gatedepend}c).

 \subsection{Temperature dependence}

\begin{figure}[ht!]
	\includegraphics[width=\columnwidth]{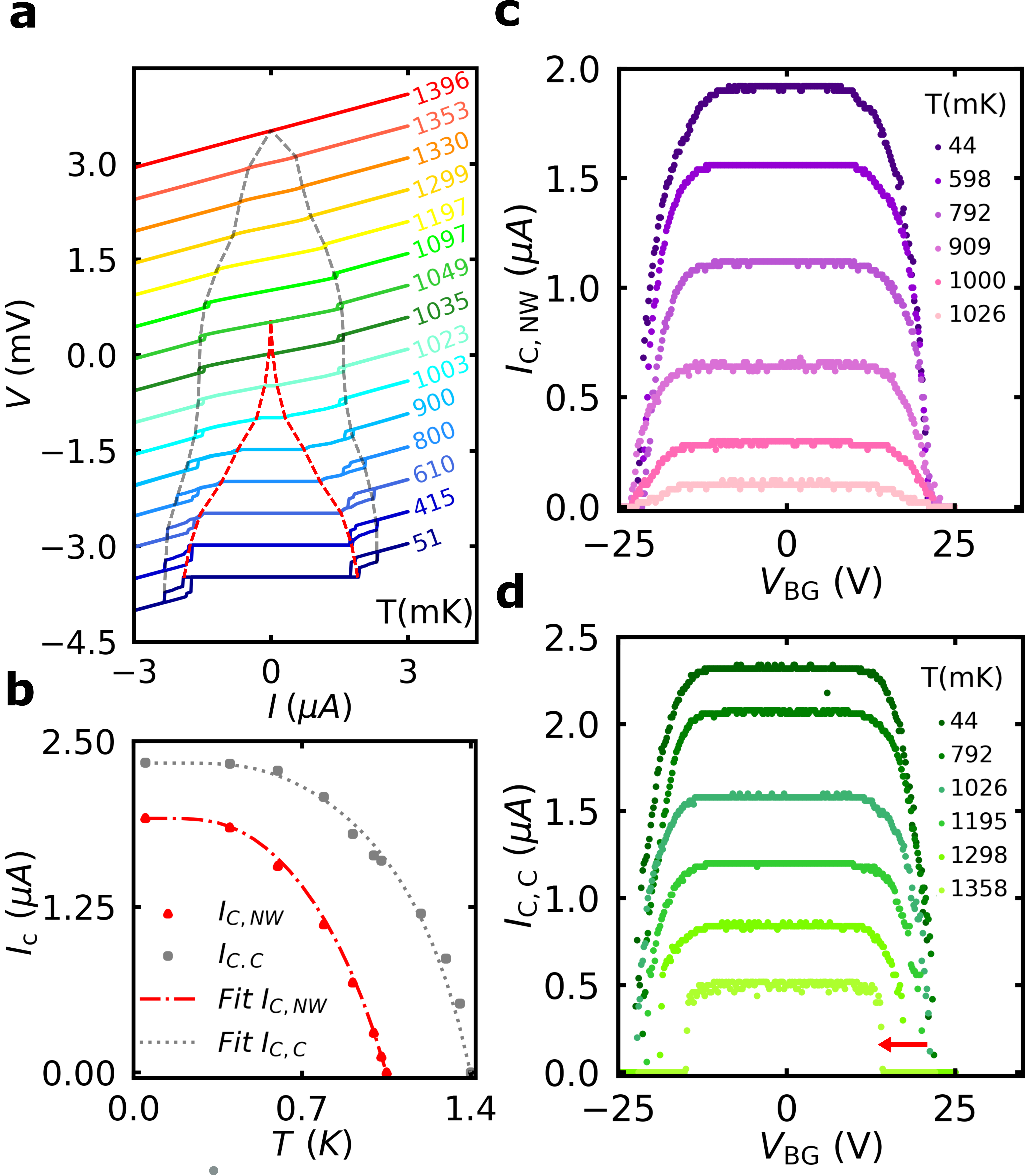}
	\caption{\label{fig:Tempdepend}\textbf{Temperature dependence} \textbf{a} (I-V) characteristics of nanowire device at elevated temperatures up to 1400 mK. Red and grey dashed lines traces the suppression of both $I_{\mathrm{C,NW}}$ and $I_{\mathrm{C,C}}$ with increasing temperature,respectively. $I_{\mathrm{C,NW}}$ switches to normal state at critical temperature $T_{\mathrm{C,NW}}$ = 1050 mK, while $I_{\mathrm{C,C}}$ switches at $T_{\mathrm{C,C}}$ = 1400 mK. \textbf{b} Temperature dependence of $I_{\mathrm{C,NW}}$ and $I_{\mathrm{C,C}}$ estimated from Fig.~\ref{fig:Tempdepend}a. $I_{\mathrm{C,NW}}$ and $I_{\mathrm{C,C}}$ extracted from panel a are fitted by using Ambegaokar Baratoff relation illustrated by red dashed-dotted and grey dotted lines, respectively. \textbf{c},\textbf{d} critical current as a function of bipolar gate voltage for both $I_{\mathrm{C,NW}}$ and $I_{\mathrm{C,C}}$ up to 98\% of their critical temperatures, respectively.    }
\end{figure}

The critical temperature was determined by measuring the IV characteristics of the device at different elevated temperatures at $V_{\mathrm{BG}}$ = 0 (see Fig.~\ref{fig:Tempdepend}a). Again both dashed red and grey lines traces the suppression of critical currents of nanowire segment $I_{\mathrm{C,NW}}$ and contact segment $I_{\mathrm{C,C}}$, respectively with increasing bath temperature $T$, respectively. In case of  $I_{\mathrm{C,NW}}$, it shows a transition to normal state at $T_{\mathrm{C,NW}}$ = 1050 mK while for $I_{\mathrm{C,C}}$ it switches at $T_{\mathrm{C,C}}$ = 1400 mK with corresponding normal state resistances $R_{\mathrm{n,NW}}$ = 135 $\Omega$ and $R_{\mathrm{n,C}}$= 191 $\Omega$, respectively. By extracting the values of $I_{\mathrm{C,NW}}$ and $I_{\mathrm{C,C}}$, the dependence of critical currents on temperature is plotted in Fig.~\ref{fig:Tempdepend}b. The red dashed-dotted and grey dotted curves are fits of the temperature dependence of $I_{\mathrm{C,NW}}$ and $I_{\mathrm{C,C}}$, respectively by using Ambegaokar Baratoff relation: \cite{senkpiel2020single,joo2019cooper}
\begin{equation}
I_{\mathrm{C}}R_{\mathrm{n}}=\frac{\pi}{2e}\tanh\left({\frac{\Delta(T)}{2\mathrm{k_{\mathrm{B}}}T}}\right),
\end{equation} 
where,
\begin{equation}
\Delta(T)= \Delta(0)\tanh\left( {a\sqrt{\frac{T_{\mathrm{C}}}{T}-1}}\right)
\end{equation}
is the superconducting gap at temperature T \cite{senapati2011spin,zhang2019broken}, $R_{\mathrm{n}}$ is the resistance of the normal metal, $\mathrm{k_{\mathrm{B}}}$ is Boltzmann constant. The temperature dependence of both $I_{\mathrm{C,NW}}$ and $I_{\mathrm{C,C}}$ is fitted using the coefficient a = 2 and 2.4, and $R_{\mathrm{n}}$ = 130 and 143 $\Omega$, respectively. The latter values of normal state resistances are in a good agreement with our experimental findings.  

The temperature dependence of GCS has been investigated by measuring the critical currents; $I_{\mathrm{C,NW}}$ and $I_{\mathrm{C,C}}$ as a function of bipolar gate voltage at elevated temperatures (see Fig.~\ref{fig:Tempdepend}c and d, respectively). By increasing the bath temperature, GCS of both $I_{\mathrm{C,NW}}$ and $I_{\mathrm{C,C}}$ is observed for all values of $T$ even when $T$ reaches almost 98\% of their critical temperature e.g. for $I_{\mathrm{C,NW}}$, $T$ = 1000 mK (see pink curve in Fig.~\ref{fig:Tempdepend}c) the gating characteristics remain the same. The pinch-off gate voltage $V_{\mathrm{BG0}}$ in case of $I_{\mathrm{C,NW}}$ did not change with increasing $T$ up to close to its critical temperature at $T_{\mathrm{C,NW}}$, while in case of $I_{\mathrm{C,C}}$, it shifts to lower values (indicated by red arrow) for measurements at temperatures higher than  $T_{\mathrm{C,NW}}$. Similar shift of $V_{\mathrm{BG0}}$ in case of $I_{\mathrm{C,C}}$ was observed in Ref.~\citenum{rocci2020gate}.

\subsection{Magnetic field dependence}

\begin{figure*}[ht!]
	\includegraphics[width=\textwidth]{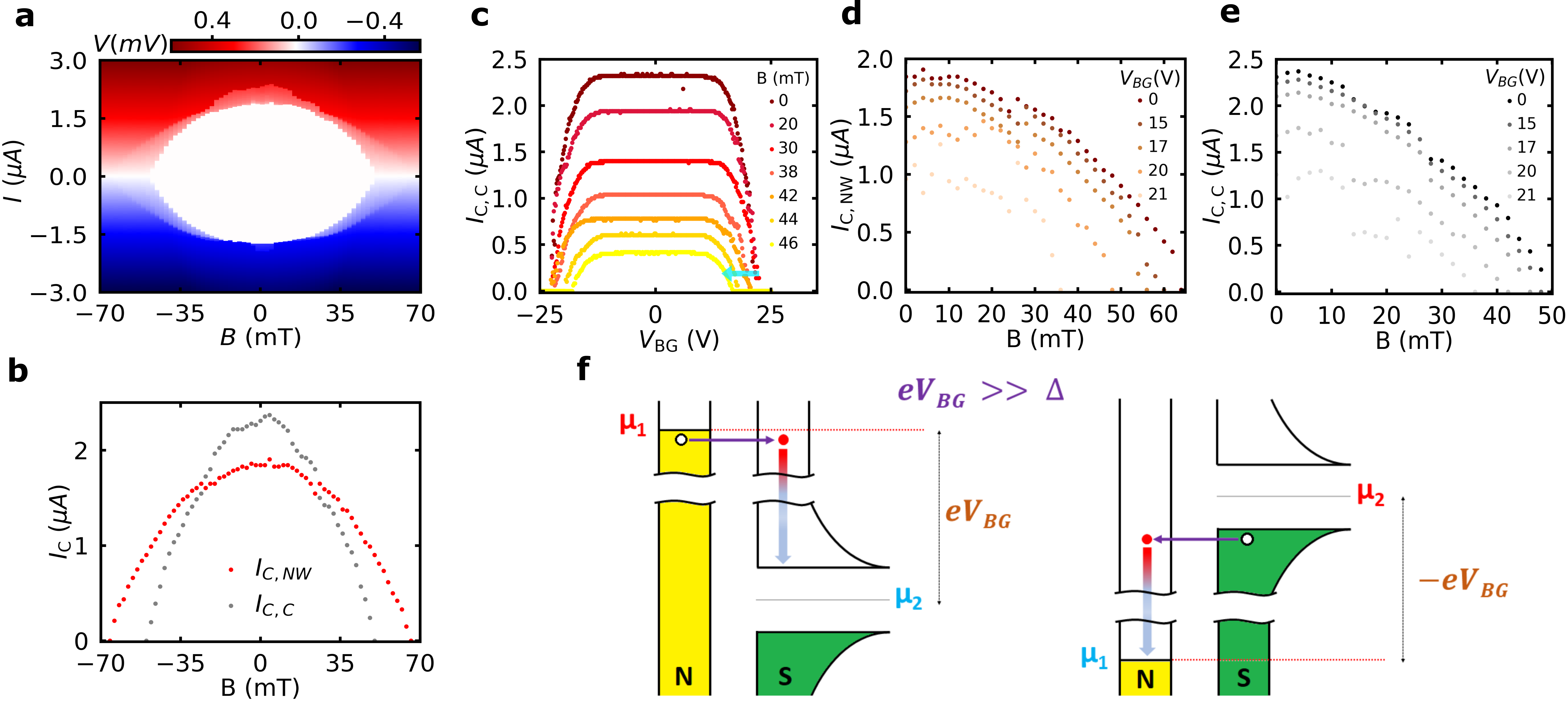}
	\caption{\label{fig:magnetdep}\textbf{Magnetic field dependence}  \textbf{a} (I-V) characteristics of nanowire device as a function of out of plane magnetic field \textbf{b} Critical currents of both nanowire device and contact interface region as a fuction of applied magnetic field extracted from measurements in panel a. $I_{\mathrm{C,NW}}$ and $I_{\mathrm{C,C}}$ crossing each others at B = ± 24 mT which results in corresponding critical magnetic fields $B_{\mathrm{C,NW}}$ = 64 mT and $B_{\mathrm{C,C}}$ = 50 mT, respectively.  \textbf{c} $I_{\mathrm{C,C}}$ as a function of bipolar gate voltage for at magnetic field values up to 46 mT. \textbf{d},\textbf{e} $I_{\mathrm{C,NW}}$ and $I_{\mathrm{C,C}}$ as a function of magnetic field at gate voltage values up to very close to pinch-off voltage, respectively. \textbf{f} Schematic diagrams of electron injection from/to metallic  gate to/from superconductor nanowire device in the left and right panel, respectively. Colored/uncolored parts represent occupied/unoccupied states, As the hot electron (red circle) tunnels it will relax to lowest unoccupied state releasing heat either in the N or in the S side, resulting different heating of the S for changing polarity of $V_{\mathrm{BG}}$.    }  
\end{figure*}

The dependence of critical currents with magnetic field was investigated by measuring (I-V) characteristics of the nanowire device as a function of out of plane magnetic field $B$ as shown in Fig.~\ref{fig:magnetdep}a. Both $I_{\mathrm{C,NW}}$ and $I_{\mathrm{C,C}}$ decreases in magnetic field, as expected. Moreover, it can be also seen that $I_{\mathrm{C,NW}}$ and $I_{\mathrm{C,C}}$ crosses each other at $B$ = ± 24 mT and their corresponding critical fields are $B_{\mathrm{C,NW}}$ = 64 mT and $B_{\mathrm{C,C}}$ = 50 mT, respectively. This is clearly seen in Fig.~\ref{fig:magnetdep}b, which shows the magnetic field dependence of critical currents extracted from measurements in  Fig.~\ref{fig:magnetdep}a.

Magnetic field dependence of GCS in nanowire device shows a similar dependence as its temperature dependence: that the gate dependence has the same general trend at all magnetic fields up to 46 mT as shown in Fig.~\ref{fig:magnetdep}c in case of $I_{\mathrm{C,C}}$. On the other hand, $V_{\mathrm{BG0}}$ decreases when $B$ values approaches to $B_{\mathrm{C,C}}$. In the same way, $I_{\mathrm{C,NW}}$ shows a similar magnetic field dependence (see supporting information). We have also measured both critical currents $I_{\mathrm{C,NW}}$ and $I_{\mathrm{C,C}}$ as a function of the magnetic field at different gate voltages up to values very close to $V_{\mathrm{BG0}}$. A significant shift in $B_{\mathrm{C,NW}}$ and $B_{\mathrm{C,C}}$ to smaller values is observed by increasing the $V_{\mathrm{BG}}$ higher than 15 V as illustrated in Fig.~\ref{fig:magnetdep} d,e, respectively. Similar dependence with B field was observed in Ti-based superconductor nanostructures \cite{de2018metallic,paolucci2019magnetotransport}. 
\section*{Discussion}

As we mentioned before, the origin of GCS was attributed in the previous studies to two different mechanisms, either to distortion of the superconducting state by induced electric field \cite{de2018metallic,de2019josephson,paolucci2019field,de2020niobium,puglia2020electrostatic,puglia2020vanadium,rocci2020gate,paolucci2019magnetotransport,paolucci2019field2,alegria2021high,ritter2021superconducting,mercaldo2020spectroscopic,mercaldo2020electrically,chirolli2020impact,solinas2021sauter}  or to high energy quasiparticle injection via tunneling \cite{alegria2021high,ritter2021superconducting,golokolenov2021origin}. We will compare our experimental findings with these explanations in the following. 

GCS was reported in various device geometries of evaporated policrystaline metallic nanobridges from different superconductors \cite{de2018metallic,de2019josephson,paolucci2019field,de2020niobium,puglia2020electrostatic,puglia2020vanadium,rocci2020gate,paolucci2019magnetotransport,paolucci2019field2,alegria2021high, ritter2021superconducting} and explanation was proposed based on electric field induced distortion of the superconducitng wave function  which could destroy the BCS state \cite{mercaldo2020spectroscopic,mercaldo2020electrically,chirolli2020impact} like e.g. by Swinger effect \cite{solinas2021sauter}. The observed $B$ field and $T$ dependencies of gating effect shows very similar characteristics to our epitaxial superconductor case. 
However, there is a finite leakage current of 100 pA at gate voltages where supercurrent gets reduced (see Fig.~\ref{fig:gatedepend}a and c). This leakage current is B-field and temperature independent, as expected (see supporting information). Assuming that the leakage takes place between the gate electrode and the nanowire, hot electrons are injected into the superconducting shell with energies several orders of magnitude higher than the SC gap. These electrons can heat up the superconducting bridge and drive it to the normal state, as it is proposed by Refs.~\citenum{alegria2021high,ritter2021superconducting,golokolenov2021origin} as a microscopic origin of the gating effect. Our basic estimation (see supporting information) of induced heat transfer also suggests that hot electrons could bring the temperature of the epitaxial shell in the range of superconducting critical temperature. Instead of using silicon dioxide or other amorphous insulators, the gate electrode and the superconductor is connected by 20-30 nm thick single crystalline hBN layer in our device, which is a large bandgap insulator commonly used as a tunnel barrier in 2D electronics\cite{britnell2012electron,bretheau2017tunnelling,fu2014large}. Considering a tunnel barrier between the gate and superconductor, the heating effect resulted from hot electron injection should show a strong asymmetric dependence on the polarity of the gate voltage. For the polarity when electrons tunnel from the gate electrode to the superconductor (see Fig.~\ref{fig:magnetdep}f left) hot electrons relax their energy in the superconductor by inducing a large number of quasiparticles, which results in a significant heat load. On the other hand, for opposite polarity (see Fig.~\ref{fig:magnetdep}f right), hot electrons heat the metal block of a large gate electrode, which has a much smaller heating effect on the superconductor isolated by the gate insulator. Such gate voltage asymmetry was reported by Ref.~\citenum{alegria2021high}, however, it does not appear in our measurements (see Fig.~\ref{fig:gatedepend}a)  after the initial training period (see supplementary information), which contradicts with a simple explanation based on hot electron injection. Moreover, very recent studies based on suspended nanobridges \cite{rocci2020gate}, where the leakage current is suppressed by several orders of magnitude, superconducting gating is also present. This supports that besides leakage another origin of gating could also exist. Considering the magnetic field dependence on the gate voltage (Fig.~\ref{fig:magnetdep}d,e), it is consistent with the hot electron injection, since increasing the gate voltage results in an increase in the rate of injection of high energy electrons to the nanowire segment. This will heat up the device and decreases the critical field of the superconducting wire to smaller values. Moreover, switching of the superconductor at smaller $V_{\mathrm{BG0}}$ with increasing the magnetic field shown in Fig.~\ref{fig:magnetdep}c can be attributed to decreasing of the critical temperature of the superconductor with the external magnetic field. So, a smaller gate voltage would bring the electronic temperature of the device up to the critical temperature of the superconductor. Finally, let us focus on the T dependence of $V_{\mathrm{BG0}}$. For temperatures larger than $T_{\mathrm{C,NW}}$ the nanowire is in the normal state and only the contacts are superconducting. As shown by the red arrow on Fig.~\ref{fig:Tempdepend}d, $V_{\mathrm{BG0}}$ decreases at elevated temperatures which is consistent with joule heating induced by hot electron injected into the nanowire. At elevated temperatures, less heat load is enough to drive the contacting segment to normal state.
On the other hand, similar suppression of $V_{\mathrm{BG0}}$ with T is not observed either for $I_{\mathrm{C,NW}}$ or for $I_{\mathrm{C,C}}$ when the nanowire is superconducting (i.e. below $T_{\mathrm{C,NW}}$), as seen on Fig.~\ref{fig:Tempdepend}c and d, which is in contradiction with the expectation for hot electron injection as an origin of GCS.

\section*{Conclusions}
In summary, we have demonstrated the superconducting gating effect in an epitaxially grown superconducting layer for the first time. We developed a novel superconducting field effect transistor using high quality single crystalline hBN insulator layer and Al shell around an InAs nanowire as the active region, which shows a supercurrent suppression at $\simeq$ 20$\,$V. Detailed magnetic field and temperature dependent characterization allowed us to compare the experimental facts with existing scenarios of superconducting gating. Although leakage current between the gate and the nanowire was detected the sign independent gating effect suggests that simple hot electron injection does not provide a complete explanation of the observed gating. Besides the fundamental interest, our results open the way to integrate superconducting switches into novel Al/InAs-based hybrid quantum architectures.
 
\section*{Methods}
InAs nanowires with 20 nm thick Al shell and total diameter approximately
100 nm were grown by Au-assisted molecular beam epitaxy (MBE). After InAs nanowire growth with a total length of 5.5 $\mu$m, the Al shell layer was epitaxially grown around the nanowire by rotating the substrate and depositing at an angle within the MBE chamber at low temperature \cite{krogstrup2015epitaxy}.

The device was fabricated by electron beam lithography (EBL) in two separate steps. In the first step, metallic gates of Ti/Au layers with thicknesses of 5/40 nm and width of 600 nm were fabricated on an intrinsic Si wafer with a 290 nm thick oxide layer. h-BN flakes were mechanically exfoliated from the bulk onto a clean 290 nm thick $\mathrm{SiO_{2}}$/Si wafer by using adhesive tape. A polydimethylsiloxane (PDMS) polymer stamp prepared with polycarbonate (PC) layer is used to pick up the h-BN flakes at $75^{\circ}$C with a thickness of 20-30 nm which can be identified from their contrast on the Si wafer under an optical microscope. The flakes are then transferred and released on the top of the bottom gates by melting the PC layer at $120^{\circ}$C. PC layer was then dissolved by left the substrate in chloroform for 20 minutes \cite{mayer2020tuning,gurram2016spin,zomer2014fast,wang2013one}. After drying the substrate, Nanowires deposited on the top of the hBN layer and aligned perpendicular to the bottom gate by a thin glass needle with a microscopic tip controlled by a hydraulic micromanipulator along with a high magnification optical microscope. Finally, four contacts of Ti/Al with thicknesses of 10/80 nm were fabricated in the second lithography step. Before this last evaporation step, nanowires were exposed to Ar-ion plasma milling for 8 minutes at 50 W to remove any oxides on the top of Al shell \cite{chang2015hard,vaitiekenas2020zero}. 

All measurements have been done on Leiden Cryogenics CF-400 top loading cryo-free dilution refrigerator system with a base temperature of 30 mK. We have used a standard power supply with series 1 M$\Omega$ resistor to inject current through nanowire via pair of Al contacts on opposite sides of nanowire and measure the voltage across the other pair by voltage differential amplifier. Leakage current was recorded by measuring the voltage across 10 M$\Omega$ preresistor connected to the gate.

\section*{Author contributions}
 T.E., O.K. and L.I. fabricated the devices, T.E., O.K., Z. S., M.B. and G.F. performed the measurements and did the data analysis. T. K. and J. N. grew the wires. K.W. and T.T. provided the high quality hBN. All authors discussed the results and worked on the manuscript. P.M. and Sz.Cs. guided the project.

\section*{Acknowledgments}
This work has received funding from Topograph FlagERA, the SuperTop QuantERA network, SuperGate Fet Open, the FET Open AndQC, and from the OTKA FK-123894 grants.
This research was supported by the Ministry of Innovation and Technology and the National Research, Development and Innovation Office within the Quantum Information National Laboratory of Hungary and by the Quantum Technology National Excellence Program (Project Nr. 2017-1.2.1-NKP-2017-00001), by the UNKP-20-5 New National Excellence Program, the János Bolyai Research Scholarship of the Hungarian Academy of Sciences, by the Carlsberg Foundation and the Danish National Research Foundation. K.W. and T.T. acknowledge support from the Elemental Strategy Initiative conducted by the MEXT, Japan, Grant Number JPMXP0112101001,  JSPS KAKENHI Grant Number 19H05790 and JP20H00354.

\bibliography{references}

\end{document}



\title{Gate-controlled supercurrent in epitaxial Al/InAs nanowires}


\author{Tosson Elalaily}\thanks{Contributed equally to this work}
 \affiliation{Department of Physics, Budapest University of Technology and Economics and Nanoelectronics 'Momentum' Research Group of the Hungarian Academy of Sciences, Budafoki ut 8, 1111 Budapest, Hungary\\}
\affiliation{Department of Physics , Faculty of Science, Tanta University, Al-Geish St., 31527 Tanta, Gharbia, Egypt.\\}

\author{Oliv\'er K\"urt\"ossy} \thanks{Contributed equally to this work}
 \affiliation{Department of Physics, Budapest University of Technology and Economics and Nanoelectronics 'Momentum' Research Group of the Hungarian Academy of Sciences, Budafoki ut 8, 1111 Budapest, Hungary\\}

\author{Zolt\'an Scher\"ubl}
 \affiliation{Department of Physics, Budapest University of Technology and Economics and Nanoelectronics 'Momentum' Research Group of the Hungarian Academy of Sciences, Budafoki ut 8, 1111 Budapest, Hungary\\}
 \affiliation{Univ. Grenoble Alpes, CEA, Grenoble INP, IRIG, PHELIQS, 38000 Grenoble, France\\}
 
\author{Martin Berke}
 \affiliation{Department of Physics, Budapest University of Technology and Economics and Nanoelectronics 'Momentum' Research Group of the Hungarian Academy of Sciences, Budafoki ut 8, 1111 Budapest, Hungary\\}
 
 \author{Gerg\H{o} F\"ul\"op}
 \affiliation{Department of Physics, Budapest University of Technology and Economics and Nanoelectronics 'Momentum' Research Group of the Hungarian Academy of Sciences, Budafoki ut 8, 1111 Budapest, Hungary\\}

\author{Istv\'an Endre Luk\'acs}
 \affiliation{Center for Energy Research, Institute of Technical Physics and Material Science, Konkoly-Thege Mikl\'os \'ut 29-33., H-1121, Budapest, Hungary\\}

\author{Thomas Kanne}
 \affiliation{Center for Quantum Devices and Nano-Science Center, Niels Bohr Institute, University of Copenhagen, Universitetsparken 5, DK-2100, Copenhagen, Denmark\\}
 
\author{Jesper Nygård}
 \affiliation{Center for Quantum Devices and Nano-Science Center, Niels Bohr Institute, University of Copenhagen, Universitetsparken 5, DK-2100, Copenhagen, Denmark\\}

\author{Kenji Watanabe}
 \affiliation{Research Center for Functional Materials, National Institute for Material Science, 1-1 Namiki, Tsukuba, 305-0044, Japan\\}
 
 \author{Takashi Taniguchi}
 \affiliation{International Center for Materials Nanoarchitectonics, National Institute for Materials Science, 1-1 Namiki, Tsukuba 305-0044, Japan\\}
 
\author{P\'eter Makk}
 \email{makk.peter@ttk.bme.hu}
 \affiliation{Department of Physics, Budapest University of Technology and Economics and Nanoelectronics 'Momentum' Research Group of the Hungarian Academy of Sciences, Budafoki ut 8, 1111 Budapest, Hungary\\}
 
 \author{Szabolcs Csonka}
 \email{szabolcs.csonka@mono.eik.bme.hu}
 \affiliation{Department of Physics, Budapest University of Technology and Economics and Nanoelectronics 'Momentum' Research Group of the Hungarian Academy of Sciences, Budafoki ut 8, 1111 Budapest, Hungary\\}


\maketitle

\section{measurements of the other pair of connected Al contacts} 
\begin{figure}[h!]
	\includegraphics[width=.6\columnwidth]{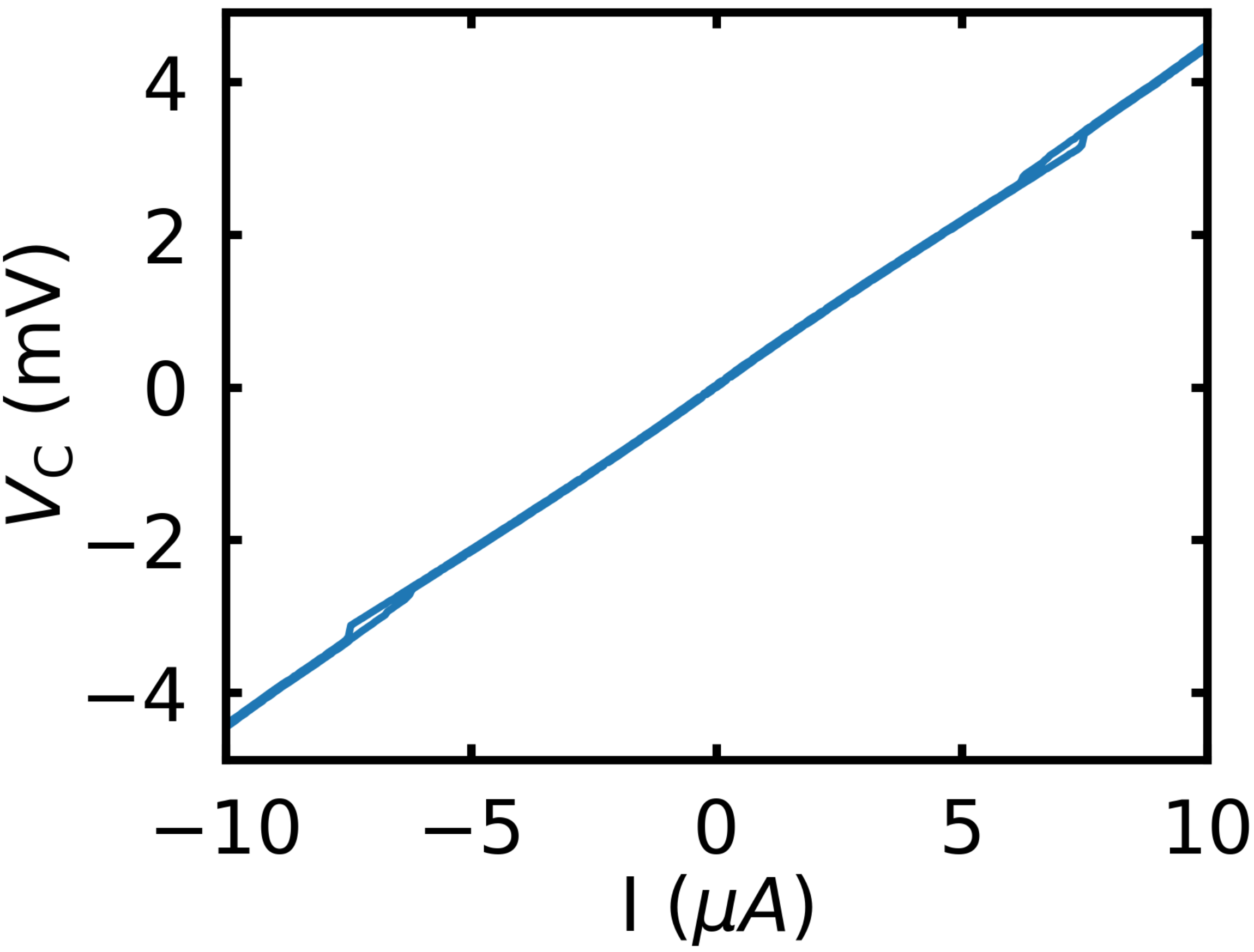}
	\caption{\label{fig:contact_measurement}\textbf{} IV characteristics of the other pair of connected Al contacts measured by 2-probe method. }
\end{figure}

In the main text, we have shown the IV characteristics of one of the contacts, here, in Fig.~\ref{fig:contact_measurement} the IV characteristics of the other pair of Al contacts measured by the 2-probe method is presented. The measurement shows switching of these contacts from the superconducting state to normal state at $\pm$ 7.5 $\mu$A which is larger than the value of both $I_{\mathrm{C,NW}}$ and $I_{\mathrm{C,C}}$ in the IV characteristics of the nanowire device in the main text.

\section{correction of the leakage current measurements}
\begin{figure}[h!]
	\includegraphics[width=.6\columnwidth]{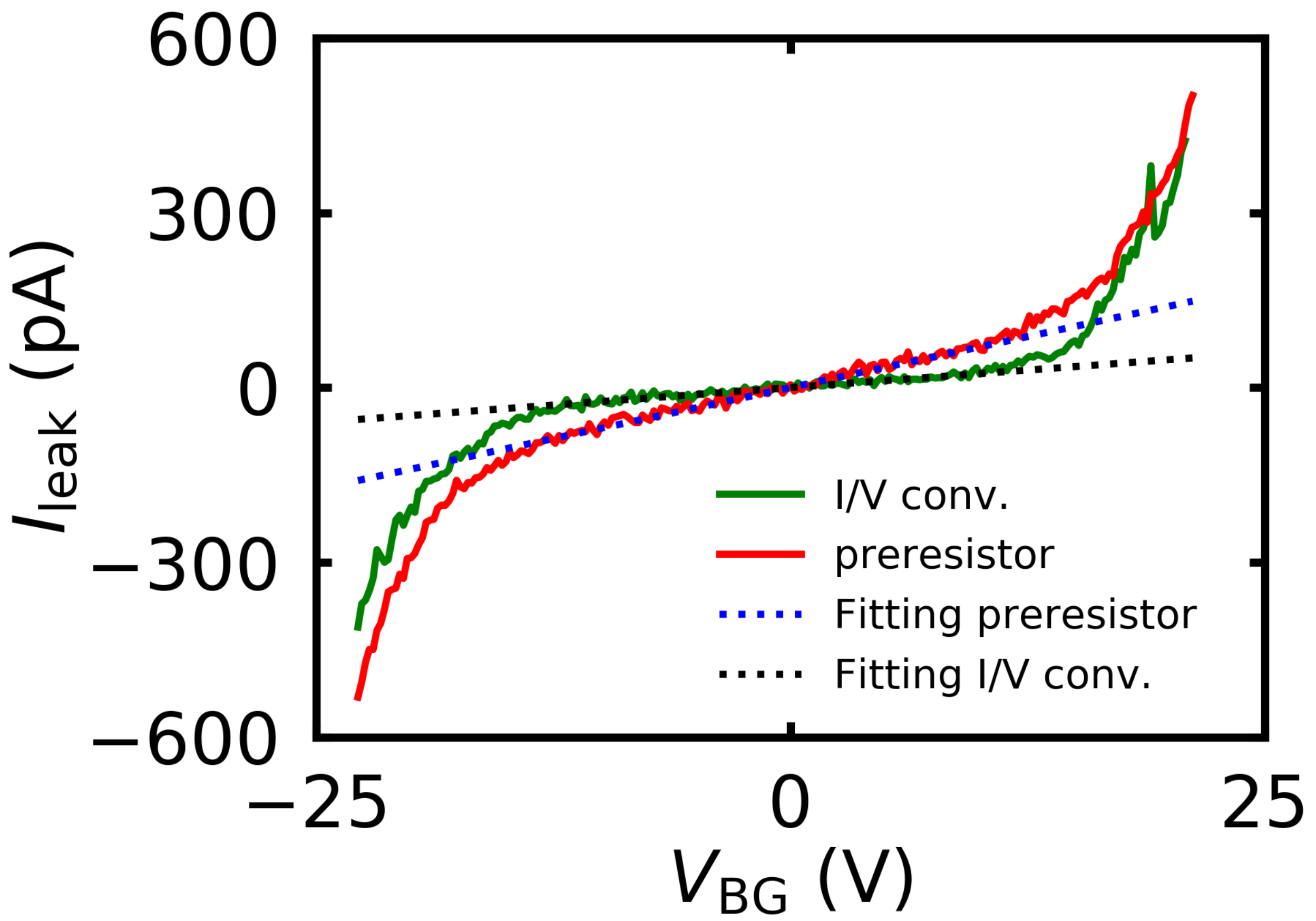}
	\caption{\label{fig:leakage_correction}\textbf{}  Leakage current measurements by preresistor (red curve) and I/V converter (green curve). The two doted lines are linear fit to the I/V converter (black) and prepresistor (blue) based leakage currents for small gate voltage values.}
\end{figure}

The leakage current measured by the preresistor connected to the gate during investigation of the device is shown in the red curve in Fig.~\ref{fig:leakage_correction} and contains two current components: a leakage current from the gate to the nanowire device plus other leakage components e.g. through the wiring of the measurement setup. To extract the exact leakage current of the investigated device, the later component should be subtracted from the measurements in the entire gate voltage range. For that purpose, we have connected one contact of the device to an I/V converter and let the other contacts on float. By sweeping the gate voltage up to $\pm V_{\mathrm{BG0}}$, the current measured through the I/V converter would give only the leakage current component to the nanowire device (green curve) \cite{ritter2021superconducting}. As Fig.~\ref{fig:leakage_correction} shows there is a difference between the current leaking from the gate (preresistor) and the current drained by the superconducting nanowire (I/V conv). Since the later current can not be measured during supercurrent measurement we estimated the leakage current to the superconducting nanowire in the following way: from the leakage current measured by the preresistor the difference between the linear slope of the preresistor measurement (blue dotted line) and I/V converter measurement  (black dotted line) was subtracted. These leakage current values are plotted in Fig. 2c in the main text.

\section{Magnetic field dependence of nanowire segment }
\begin{figure}[h!]
	\includegraphics[width=.6\columnwidth]{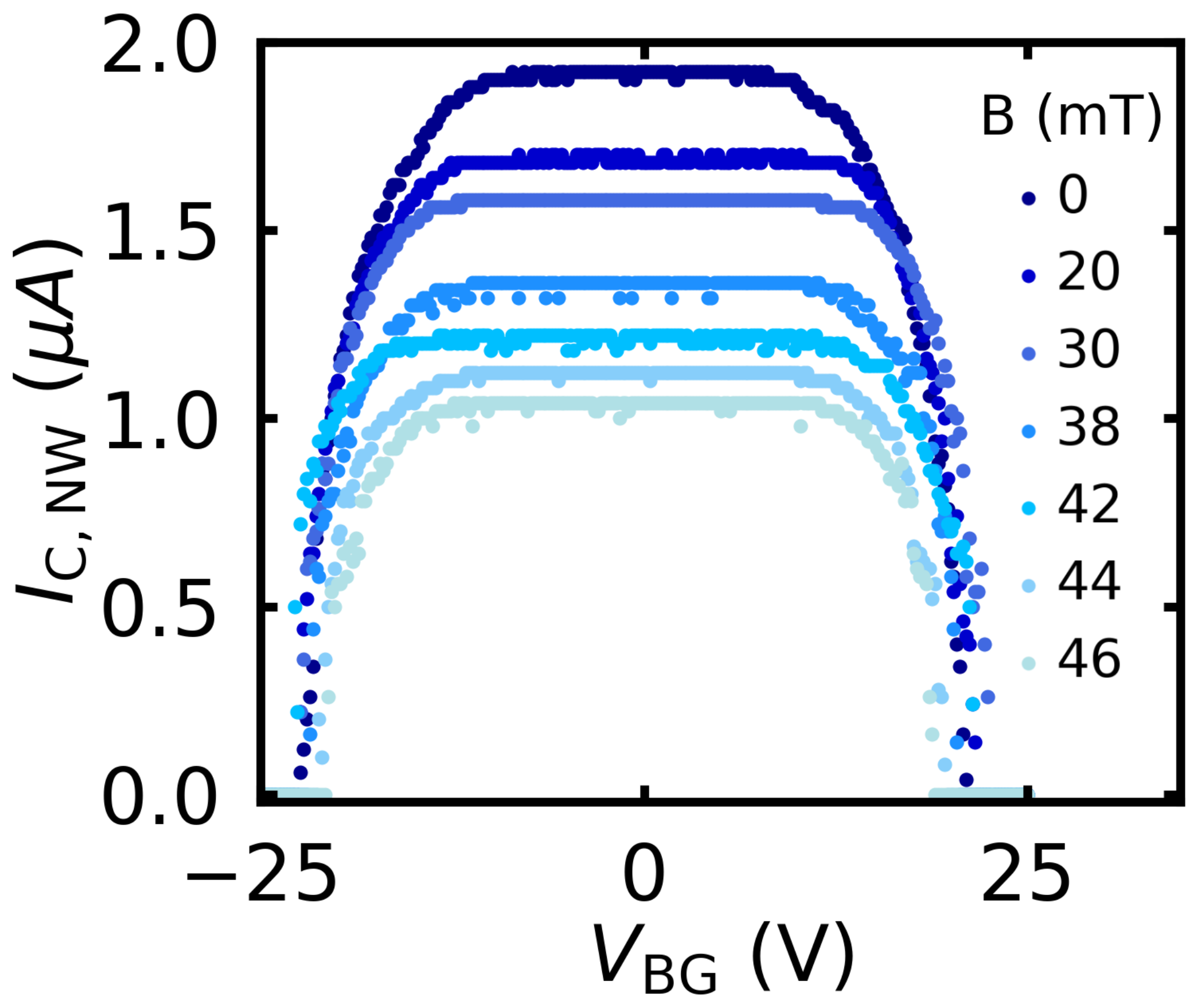}
	\caption{\label{fig:magdepend}\textbf{}  Magnetic field dependence of critical current of nanowire segment $I_{\mathrm{C,NW}}$. }
\end{figure}

Fig.~\ref{fig:magdepend} shows the dependence of the critical current of nanowire segment $I_{\mathrm{C,NW}}$ under influence of the external magnetic field B. Since the observed dependence of $V_{\mathrm{BG0}}$ on B is strong at field values very close to the critical field, the magnetic field dependence of $I_{\mathrm{C,NW}}$ show a weaker dependence compared to that of the contact segment $I_{\mathrm{C,C}}$ (Fig.4c in the main text) because the measurements are carried out only up to 46 mT which is a bit far from the critical field value of nanowire segment $B_{\mathrm{C,NW}}$ = 64 mT, while for the contact segment $B_{\mathrm{C,C}}$ = 50 mT.

\section{Influence of temperature and magnetic field on the leakage current}
\begin{figure}[h!]
	\includegraphics[width=\columnwidth]{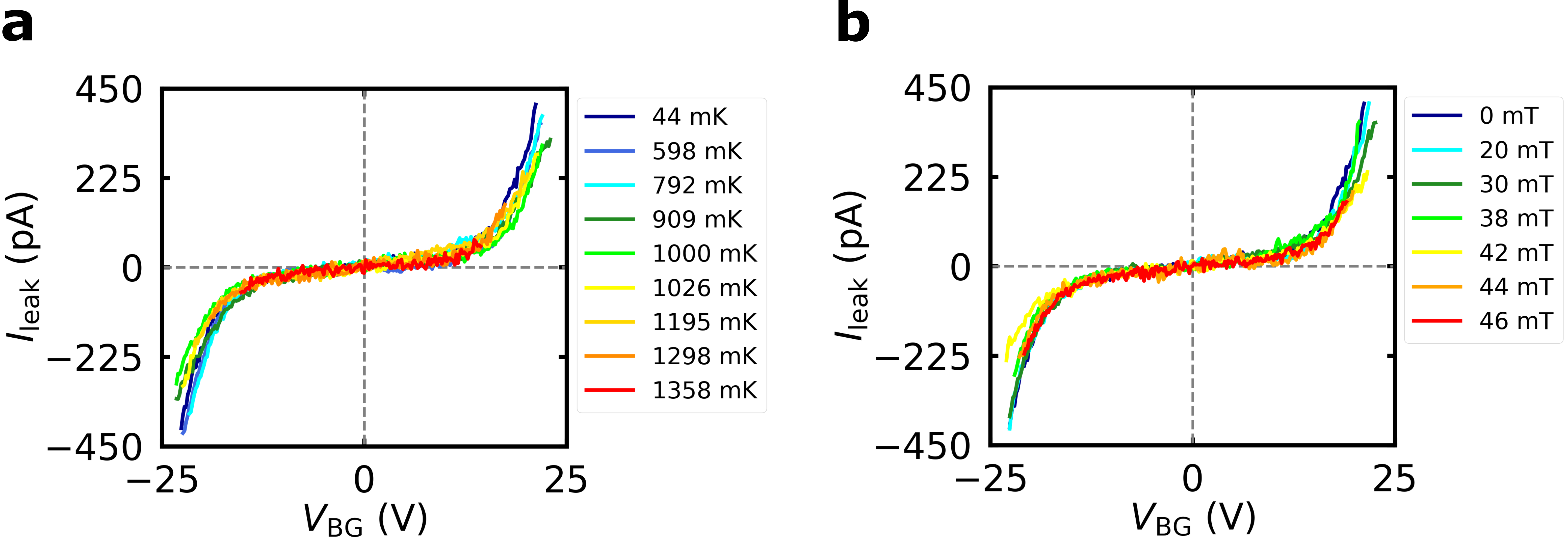}
	\caption{\label{fig:leakage_dependence}\textbf{} \textbf{(a)} The leakage current $I_{\mathrm{leak}}$ as a function of the gate voltage $V_{\mathrm{BG}}$ at elevated temperatures and \textbf{(b)} at different values of external magnetic field. }
\end{figure}

The temperature and magnetic field dependencies of the leakage current measured during investigation of the device are shown in Fig~\ref{fig:leakage_dependence}a,b, respectively after correction with the method explained in section II. The measurements show that the measured leakage current is independent of either the elevated temperatures or the external magnetic field. 

\section{Thermal model for calculating the electronic temperature of nanowire device. }
\begin{figure}[h!]
	\includegraphics[width=\columnwidth]{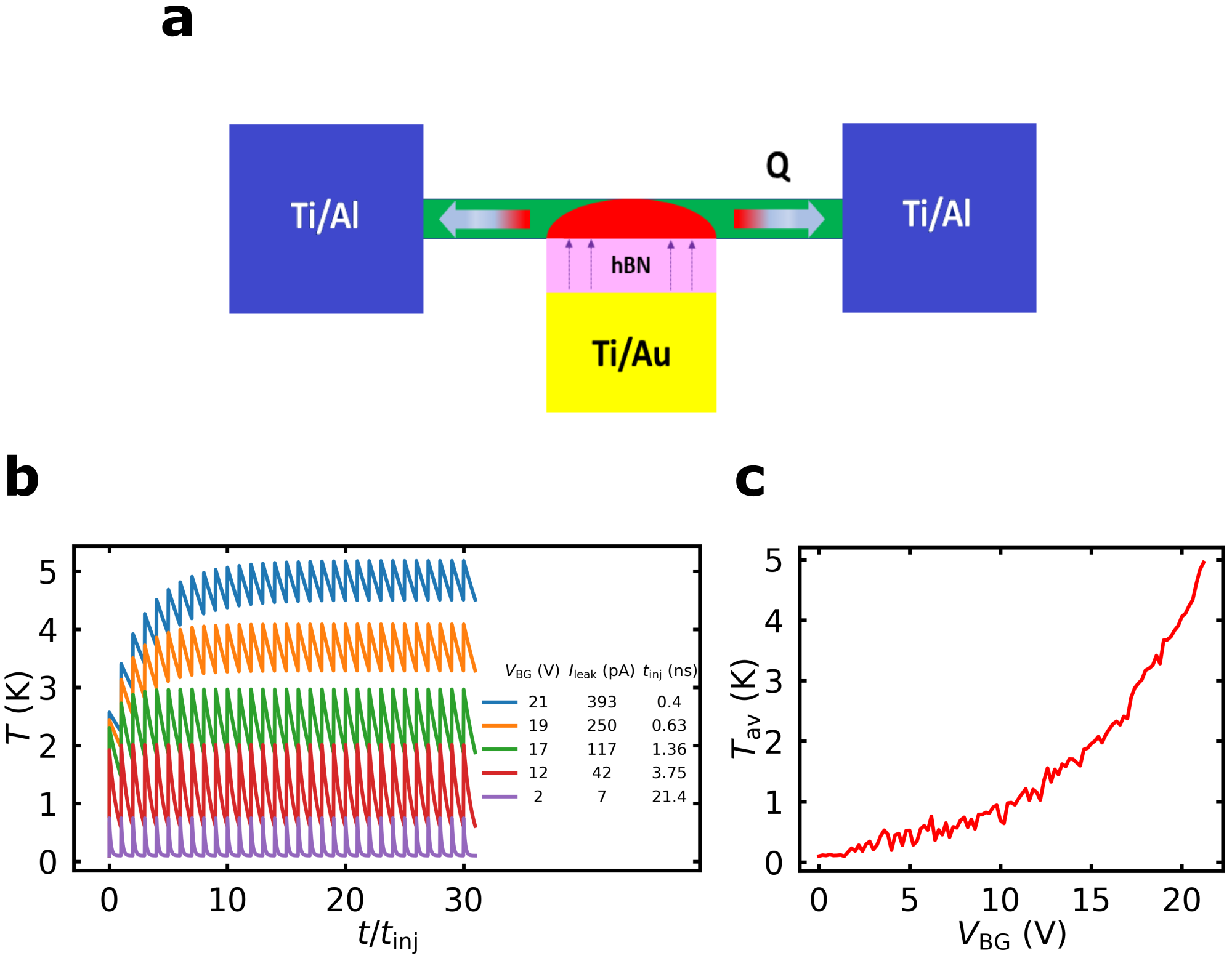}
	\caption{\label{fig:thermal model}\textbf{} \textbf{(a)} Schematic of the thermal model for calculating the electronic temperature of nanowire device after injection of of electrons with energy $eV_{\mathrm{g}}$ from the gate electrode. \textbf{(b)} Calculated electronic temperature of the nanowire segment within 30 successive electron injection cycles at different gate voltages, where $t_{\mathrm{inj}}$ = $e/I_{\mathrm{leak}}$.\textbf{(c)} The average nanowire temperature over 20 injection cycles as a function of gate voltage.}
\end{figure}

In this section, we will give a simple estimation for the increase of the electronic temperature of the nanowire segment resulting from the injection of hot electrons from the gate. Suppose that a single electron tunnels from the gate electrode to the Al layer (that has a volume $\Omega$) with energy $eV_{\mathrm{BG}}$, that is much higher than the superconducting gap of Al as illustrated in the schematic in Fig~\ref{fig:thermal model}a. This electron will release its energy to the electron gas. Assuming that the temperature of the Al shell is initially at temperature $T_{\mathrm{i}}$, the injection of an electron to the Al shell will result in a sharp increase on its electronic temperature derived from the heat equation: \cite{rocci2020gate,puglia2020electrostatic} 
\begin{equation}
T_{i+1}=\sqrt{\frac{2eV_{\mathrm{BG}}}{\Omega\gamma}+T_{\mathrm{i}}^2},
\end{equation}
where $\gamma$ is the Sommerfeld coefficient. The temperature difference between the nanowire segment and the 
contacts which are at base temperature $T_{\mathrm{b}}$ = 100 mK will result in a cool-down of the nanowire segment through heat diffusion by electrons to the contacts.
The amount of heat transferred within time $dt$ is given by:  
\begin{equation}
Q=\frac{KA\Delta T}{L}\cdot dt,
\end{equation}
where the Wiedemann-Franz law is used:
\begin{equation}
K= \sigma \mathcal{L} T_{i},
\end{equation}
where K is the thermal conductivity of the Al shell, $\sigma$ = $R_{\mathrm{n}}L$/$A$ is its electrical conductivity, $L$ is the half length of the nanowire segment from the injection point to the contacts, $A$ is its cross-sectional area, $R_{n}$ is its normal state resistance, $\mathcal{L}$ is the Lorentz number and $\Delta T$ is the temperature difference between actual temperature of the middle of the nanowire, $T_{i}$ and the temperature of the contact $T_{\mathrm{b}}$: $\Delta T$ =$T_{\mathrm{i}}$ - $T_{\mathrm{b}}$. Due to the heat loss to the contacts, the electronic temperature of the nanowire segment $T_{i}$ will be decreased to $T_{i+1}$ given by the following heat loss equation:
\begin{equation}
T_{i+1}=T_{i}-\frac{Q}{\Omega \gamma T_{i}}\cdot
\end{equation}
Assuming that at every time period of $t_{\mathrm{inj}}$ = $e/I_{\mathrm{leak}}$ a hot electron arrives and in the meantime the wire continuously cools, the time dependent temperature can be simulated  by Eqs. 1 - 4 in discretized points $T_{\mathrm{i}}$, $i$=1, 2, 3, .... 

Fig~\ref{fig:thermal model}b shows the calculated temperature as a function of time, for a time period that contains 30 successive electron injection cycles ($t_{inj}$) at different gate voltages using $\Omega$ = $7.536 \times 10^{-21}$ $\mathrm{m^3}$ ,$\gamma$ = 135.1 $\mathrm{J m^{-3} K^{-2}}$, $R_{\mathrm{n}}$= 135 $\Omega$, $A$= $5.024 \times 10^{-15}$ and $\mathcal{L}$= $2.45 \times 10^{-8}$ $V^2$ $K^{-2}$. At smaller gate voltages, the leakage current is quite small and the corresponding electron injection time $t_{\mathrm{inj}}$ is very long (see table in Fig~\ref{fig:thermal model}b), the nanowire segment will cool down to base temperature before the next electron is injected (see purple curve in Fig~\ref{fig:thermal model}b), while at larger gate voltages, $t_{\mathrm{inj}}$ is very small and the system could not cool down close to base temperature before the next electron is injected. In the latter case, the successive injections will increase the effective temperature of the nanowire until it reaches a steady state temperature (see blue curve Fig~\ref{fig:thermal model}b). Fig~\ref{fig:thermal model}c shows the average nanowire temperature (taken over 20 injection cycles at the saturation region) as a function of gate voltage. The calculated temperature increases exponentially with increasing gate voltage up to values in the same order of magnitude as the critical temperatures $T_{\mathrm{C,NW}}$ = 1 K and $T_{\mathrm{C,C}}$ = 1.4 K. The above calculation is based on some simplifications, like the heat transfer via phonons is neglected, the wire is handled  as a normal conductor. Based on the above simple estimation, leakage current-based suppression of the supercurrent could be a feasible explanation of superconducting gating, since the heat transfer of hot electrons could bring the superconducting nanobridge to normal state.
 
\section{Initial training  measurements on nanowire device}
\begin{figure}[H]
	\includegraphics[width=\columnwidth]{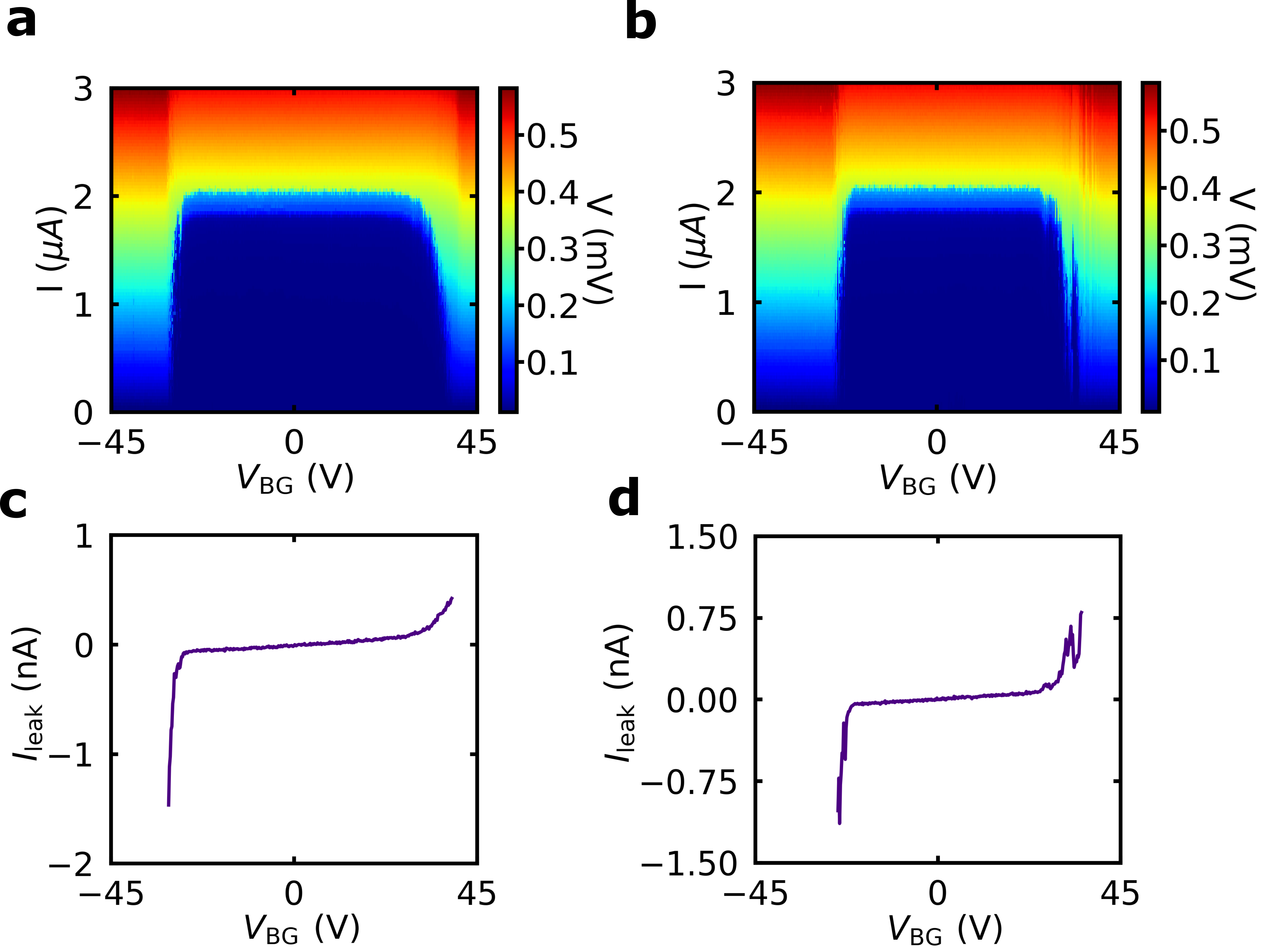}
	\caption{\label{fig:earlier test}\textbf{} \textbf{(a)},\textbf{(b)} IV characteristics of nanowire device as a function of bipolar gate voltage $V_{\mathrm{BG}}$ for two successive measurements and thier corresponding leakage current in \textbf{(c)},\textbf{(d)}, respectively. }
\end{figure}

Investigation of the gating effect at the earlier training measurements on the nanowire device shows asymmetry on the supercurrent-gate voltage dependence and the device switches to normal state at higher gate voltage (see Fig~\ref{fig:earlier test}a) compared to symmetric measurements in Fig.2a in the main text. The asymmetry on the supercurrent dependence is accompanied by a similar asymmetry on the leakage current recorded during the measurements as shown in Fig~\ref{fig:earlier test}c. By repeating the measurements, the supercurrent dependence and the corresponding leakage current becomes more symmetric on the bipolar gate voltage (see Fig~\ref{fig:earlier test} b,d, respectively) and the pinch-off gate voltage $V_{\mathrm{BG0}}$ is decreased to lower values. Ramping the gate voltage back and forth many times leads to symmetric measurements shown in the main text. 
\bibliography{references}